\documentclass[pdflatex,sn-mathphys-num]{sn-jnl}

\usepackage{graphicx}%
\usepackage{multirow}%
\usepackage{amsmath,amssymb,amsfonts}%
\usepackage{amsthm}%
\usepackage{mathrsfs}%
\usepackage[title]{appendix}%
\usepackage{xcolor}%
\usepackage{textcomp}%
\usepackage{manyfoot}%
\usepackage{booktabs}%
\usepackage{algorithm}%
\usepackage{algorithmicx}%
\usepackage{algpseudocode}%
\usepackage{listings}%
\usepackage[numbers]{natbib}

\theoremstyle{thmstyleone}%

\theoremstyle{thmstyletwo}%

\theoremstyle{thmstylethree}%

\begin{document}

\title[Article Title]{Large model enhanced computational ghost imaging}


\author[1,2]{\fnm{Yifan} \sur{Chen}}
\equalcont{These authors contributed equally to this work.}

\author[1,2]{\fnm{Hongjun} \sur{An}}
\equalcont{These authors contributed equally to this work.}

\author*[1,2]{\fnm{Zhe} \sur{Sun}}\email{sunzhe@nwpu.edu.cn}
\equalcont{These authors contributed equally to this work.}
\author[3,4]{\fnm{Tong} \sur{Tian}}
\author[5,6]{\fnm{Mingliang} \sur{Chen}}
\author[3,4]{\fnm{Christian} \sur{Spielmann}}
\author*[1,2]{\fnm{Xuelong} \sur{Li}}\email{li@nwpu.edu.cn}

\affil[1]{\orgdiv{School of Artificial Intelligence, OPtics and ElectroNics (iOPEN)}, \orgname{Northwestern Polytechnical University}, \orgaddress{ \city{Xi’an}, \postcode{710072}, \country{ P. R. China}}}

\affil[2]{\orgdiv{Institute of Artificial Intelligence (TeleAI)}, \orgname{China Telecom}, \orgaddress{\city{Shanghai}, \postcode{200232}, \country{P. R. China}}}

\affil[3]{\orgdiv{Institute of Optics and Quantum Electronics, Abbe Center of Photonics}, \orgname{Friedrich Schiller University}, \orgaddress{ \city{Jena}, \postcode{07743},  \country{Germany}}}

\affil[4]{\orgname{Helmholtz Institute Jena}, \orgaddress{ \city{Jena}, \postcode{07743},  \country{Germany}}}

\affil[5]{\orgdiv{Aerospace Laser Technology and System Department, Shanghai Institute of Optics and Fine Mechanics}, \orgname{Chinese Academy of Sciences}, \orgaddress{ \city{ Shanghai}, \postcode{201800},  \country{P. R. China}}}

\affil[6]{\orgdiv{Center of Materials Science and Optoelectronics Engineering}, \orgname{University of Chinese Academy of Sciences}, \orgaddress{ \city{Beijing}, \postcode{100049},  \country{P. R. China}}}

\abstract{Ghost imaging (GI) achieves 2D image reconstruction through high-order correlation of 1D bucket signals and 2D light field information, particularly demonstrating enhanced detection sensitivity and high-quality image reconstruction via efficient photon collection in scattering media. Recent investigations have established that deep learning (DL) can substantially enhance the ghost imaging reconstruction quality. Furthermore, with the emergence of large models like SDXL, GPT-4, etc., the constraints of conventional DL in parameters and architecture have been transcended, enabling models to comprehensively explore relationships among all distinct positions within feature sequences. This paradigm shift has significantly advanced the capability of DL in restoring severely degraded and low-resolution imagery, making it particularly advantageous for noise-robust image reconstruction in GI applications. In this paper, we propose the first large imaging model with 1.4 billion parameters that incorporates the physical principles of GI (GILM). The proposed GILM implements a skip connection mechanism to mitigate gradient explosion challenges inherent in deep architectures, ensuring sufficient parametric capacity to capture intricate correlations among object single-pixel measurements. Moreover, GILM leverages multi-head attention mechanism to learn spatial dependencies across pixel points during image reconstruction, facilitating the extraction of comprehensive object information for subsequent reconstruction. We validated the effectiveness of GILM through a series of experiments, including simulated object imaging, imaging objects in free space, and imaging object located 52 meters away in underwater environment. The experimental results show that GILM effectively analyzes the fluctuation trends of the collected signals, thereby optimizing the recovery of the object’s image from the acquired data. Finally, we successfully deployed GILM on a portable computing platform, demonstrating its feasibility for practical engineering applications.}




\maketitle

\section{Introduction}\label{sec1}

Ghost imaging (GI) is a non-local computational imaging approach that reconstructs objects by correlating the spatially modulated illumination patterns and temporally resolved single-pixel intensity fluctuations. This quantum-inspired methodology exploits statistical fluctuations in photon counting sequences, offering distinctive advantages for high-fidelity image reconstruction in photon-limited environments and highly scattering media. Building on these principles, GI has found broad applicability in fields such as microscopic imaging \cite{1,2}, X-ray imaging \cite{3,4}, and LiDAR imaging \cite{5,6}. The foundational implementation by Shih et al. \cite{7,8} utilized biphoton quantum entanglement, establishing the quantum optical framework for GI. However, the practical deployment of quantum ghost imaging (QGI) remains constrained by wavelength-specific entanglement generation challenges \cite{8} and inherent flux limitations. In contrast, Bennink et al. \cite{9} demonstrated classical GI, overcoming quantum constraints by using classical source, albeit with a cost of reduced Signal-to-Noise Ratio. Subsequently, Valencia et al. \cite{10} achieved GI using a pseudo-thermal light source. To further simplify the experimental setup and enhance the controllability of the light field, subsequent innovations by Shapiro \cite{11} introduced computational GI (CGI) using programmable spatial light modulator (SLM), which simplified the optical architecture to single-arm configurations while enabling precise wavefront control. Since CGI also uses a pseudo-thermal light source, it is essentially a type of pseudo-thermal ghost imaging (PGI). Therefore, the visibility of the images reconstructed by CGI is theoretically limited to $33\%$ of that in QGI, due to the inherent incoherence of the pseudo-thermal light source \cite{12,13}. Moreover, CGI requires extensive single-pixel sampling, as each sample provides only limited information about the object. Consequently, improving sampling efficiency and enhancing image contrast have become central challenges in CGI, driving research into compressive sensing algorithms and nonlinear correlation extraction techniques to address both issues simultaneously.

Compressed sensing \cite{14} capitalizes on the sparsity assumption of signals to effectively compress data during the acquisition, and has been integrated into CGI to reduce sampling requirements. Duarte et al. \cite{15} applied compressive sensing to computational imaging, enabling high-quality reconstruction of object images from sub-Nyquist sampling rate. Katz et al. \cite{16} further advanced this approach by developing an image reconstruction algorithm for PGI, capable of reconstructing N-pixel images from fewer than N measurements. Compressive sensing-based GI (GICS) methods have proven effective in reducing data acquisition, minimizing noise, and enhancing system flexibility \cite{17,18,19}. However, although compressed sensing can significantly reduce the number of samples required in CGI, it still struggles to substantially improve the visibility of the reconstructed images.

Recently, with the development of deep learning (DL) and its remarkable success in image enhancement \cite{20,21}, image recognition \cite{22,23}, and segmentation \cite{24,25}, there has been a concerted effort to incorporate DL into CGI \cite{26,27,28}. Lyu et al. \cite{29} proposed GI using DL (GIDL), where neural networks are trained to map images reconstructed by traditional GI techniques and their corresponding ground truth. The trained neural network then improves the reconstruction quality, enabling high-quality imaging at extremely low sampling rates. Shimobaba et al. \cite{30} developed a deep neural network to automatically learn features of noise-contaminated images reconstructed by computational GI, thus greatly improving the image reconstruction quality. Yang et al. \cite{31} utilized a generative adversarial network (GAN) to enhance the object reconstruction in underwater GI, while Zhang et al. \cite{32} used multi-scale generative adversarial network (MsGAN) to optimize GI performance in turbulent water. Although the aforementioned methods can significantly improve the image reconstruction quality, they often rely on pre-trained models for image background denoising, which requires large datasets. This reliance increases the cost of dataset acquisition and limits model performance in scenarios where these methods have not been pre-trained.

To reduce dataset acquisition costs and enhance model generalization across different scenarios, self-supervised methods \cite{33,34} have been introduced to train models without the need for labeled data. Wang et al. \cite{35,36} integrated a physical model required for GI into a neural network, using light intensity signals from the physical setup to impose constraints on the network’s output. Our previous work \cite{37} proposed a self-supervised information extraction network for underwater GI, which speckle information extraction module is designed to improve the utilization of speckle information and object information extraction module is used to extract object information from the image generated by GI. In another work \cite{38}, we proposed a part-based self-supervised image-loop neural network for single pixel imaging, incorporating prior information by looping self-reconstructed image feedback to the network refinement. Li et al. \cite{39} and Zhang et al. \cite{40} employed untrained networks to reconstruct 2D image from the 1D photodiode data, incorporating physical priors to improve the image reconstruction quality. These self-supervised DL-based GI methods eliminate the need for additional data collection, thus reducing acquisition costs and improving model generalization. However, current DL-based GI methods often suffer from underfitting when reconstructing complex object images, as limitations in model parameters and architecture hinder effective mapping between single-pixel measurements and intricate object features, thus impacting image visibility. Recent advances in large-scale models, such as SDXL, GPT-4, etc. have broken conventional DL limitations by increasing model parameter capacity and improving architecture flexibility. These developments enable models to fully explore complex relationships in feature sequences, offering promising solutions for restoring severely degraded or low-resolution images. This has paved the way for more robust noise-resistant image reconstruction in GI applications. 

In this paper, we propose for the first time a large imaging model with 1.4 billion parameters, incorporating the physical model of GI into the model architecture (GILM). GILM integrates skip connection mechanisms to mitigate gradient explosion issues associated with deeper models, ensuring sufficient parameter capacity to capture the complex correlations in single-pixel measurements. Furthermore, GILM utilizes multi-head attention mechanisms to learn spatial dependencies across pixels, enhancing the extraction of comprehensive object information for reconstruction. By embedding the physical model of GI, GILM dynamically adjusts the model parameters during the imaging process, enhancing generalization capability. We have deployed GILM on a portable computing platform and validated its effectiveness through a series of experiments, including simulated object imaging, imaging objects in free space, and imaging object located 52 meters away in underwater environment.

\section{Methods}\label{sec2}

\begin{figure*}[htbp]
  \centering
  \includegraphics[width=13cm]{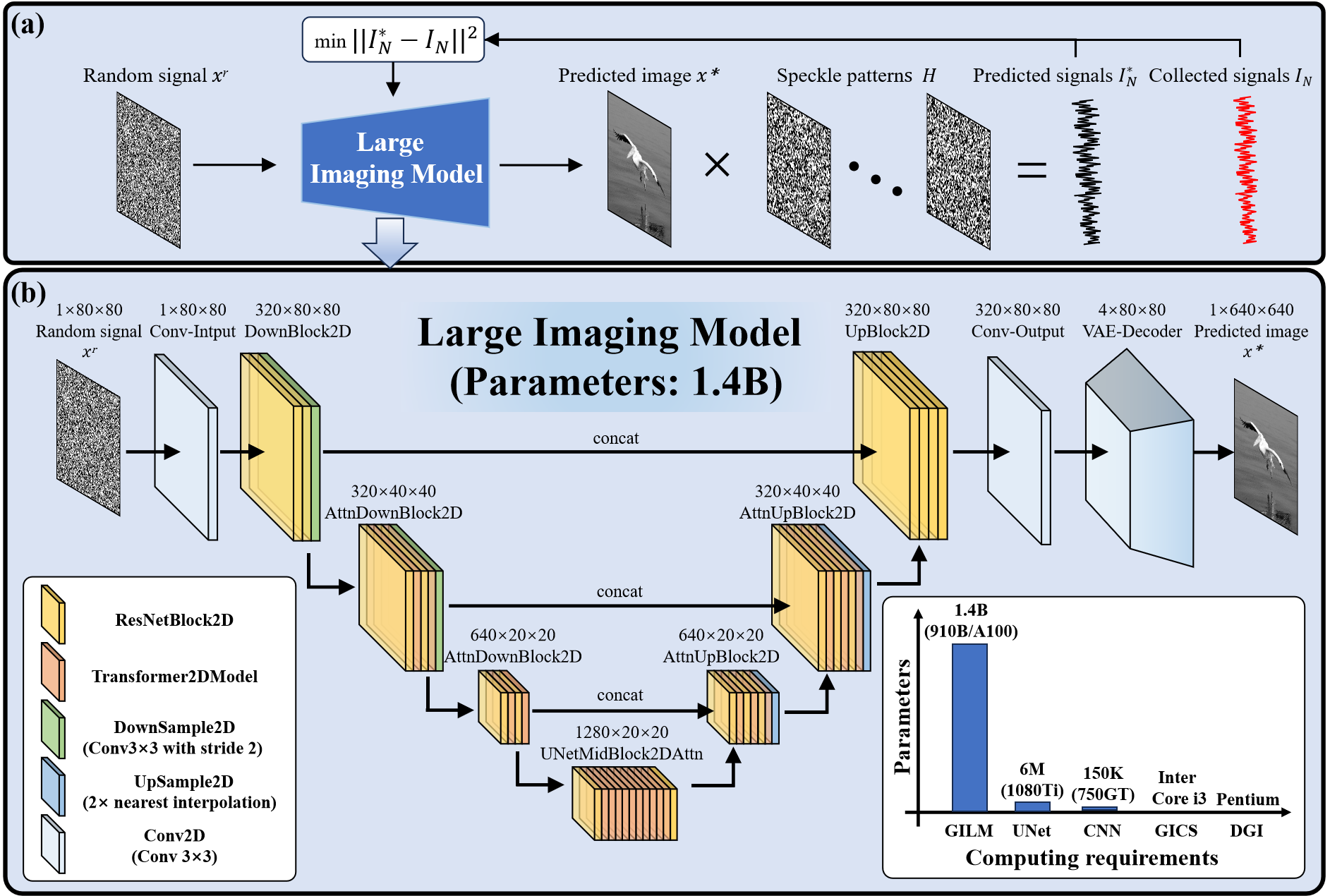}
  \caption{Overview of GILM. (a) The software framework of GILM algorithm. (b) The network structure of proposed large imaging model.
 }
  \label{fig1}
\end{figure*}

We will introduce our method from two aspects: the image reconstruction method and the specific network architecture of the proposed large imaging model. An overview of the proposed method is presented in Fig. \ref{fig1}.

\subsection{Image Reconstruction Method}
As illustrated in Fig. \ref{fig1}(a), the proposed large imaging model reconstructs an image of the object by taking a random signal $x^r$ as input. The model predicts the object image $x^*$, and by performing an element-wise multiplication of the speckle pattern $H$  and the predicted image $x^*$, the predicted intensity signals $I_N^*$ are generated. These predicted intensity signals are then compared to the collected intensity signals $I_N$ detected by the single-pixel detector (SPD). The image reconstruction process is guided by minimizing the discrepancy between the predicted and collected intensity signals, thereby enabling the model to reconstruct an image that closely approximates the true object. The reconstruction process can be formally expressed as:
\begin{align}
x^* &= f_{LIM}(x^r) \\
I_N^* &= x^* \cdot H \\
L_{\text{LIM}} &= \| I_N - I_N^* \|^2
\end{align}

where $f_{LIM}(\cdot)$ represents the large imaging model GILM, and  
$L_{LIM}$ denotes the loss function used to optimize the model’s parameters during training. This loss function constrains the convergence of the GILM’s parameters by minimizing the squared difference between the predicted intensity signals $I_N^*$ and the collected signals $I_N$. The self-supervised training method employed here utilizes the collected intensity signals as labels, allowing the model to adjust its parameters dynamically during the imaging process. This approach eliminates the need for pre-trained scene-specific data, significantly enhancing the model’s generalization capability across different imaging conditions.

\subsection{Network structure}

As can be seen from Fig. \ref{fig1}(b), the proposed large imaging model consists of several key components: ResNetBlock2D modules, Transformer2dModel modules, DownSample2D modules, UpSample2D modules, and Conv2D modules. Among these, the ResNetBlock2D module adopts a skip connection mechanism to mitigate the challenge of gradient explosion inherent in deep architectures, ensuring that the model has sufficient parameter capacity to capture the complex correlations between single-pixel measurements and the object. The Transformer2dModel module leverages multi-head attention mechanisms to learn spatial dependencies across pixels during the image reconstruction process, aiding in extraction of comprehensive object information for subsequent reconstruction. The DownSample2D module is designed to remove redundant information during the imaging process, comprising a convolutional layer with a 3×3 kernel and a stride of 2 to downsample the features while preserving essential details. The UpSample2D module increases the spatial resolution of the features to restore the desired image size, utilizing nearest-neighbor interpolation. The Conv2D module is employed for feature extraction, featuring a 3×3 kernel with a stride of 1. Furthermore, the proposed large imaging model operates within the latent space \cite{41}, enabling high-resolution image reconstruction in a relatively low-dimensional solution space. The optimal solution in the latent space is then mapped to the pixel space to form the final image, using the decoder structure of a Variational Autoencoder (VAE) \cite{42}. This design significantly reduces the computational complexity and accelerates model convergence. It is worth noting that the proposed model is the first large imaging model with 1.4 billion parameters. Preliminary estimates suggest that, before quantization and compression, it requires an A100 or 910B GPU to meet the computational requirements. The detailed descriptions of the ResNetBlock2D and Transformer2dModel modules will be provided in the following sections.

\subsubsection{The structure of ResNetBlock2D}

\begin{figure*}[htbp]
  \centering
  \includegraphics[width=12cm]{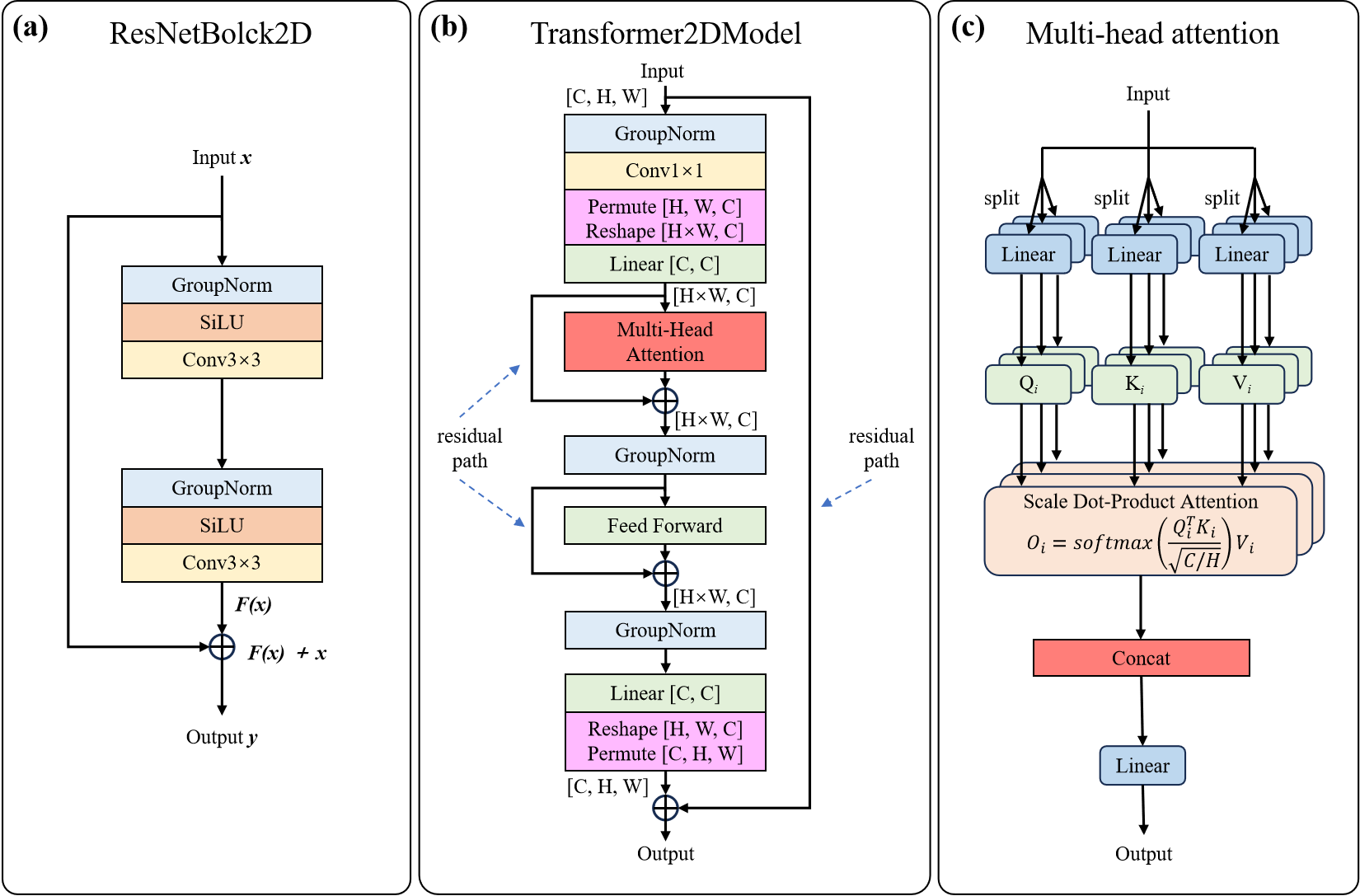}
  \caption{The network architecture of the modules in the proposed large imaging model. (a) The structure of ResNetBolck2D. (b) The structure of Transformer2DModel. (c) The structure of the multi-head attention mechanism embedded in the Transformer2DModel.
 }
  \label{fig2}
\end{figure*} 

As shown in Fig. \ref{fig2}(a), the ResNetBlock2D consists of two paths: the main path and a skip connection. The main path includes several convolutional layers, batch normalization layers, and SiLU activation functions, while the skip connection directly adds the input $x$ to the output $F(x)$ of the main path. The mathematical representation of ResNetBlock2D can be defined as:

\begin{align}
y = F(x)+x
\end{align}
The skip connection in ResNetBlock2D enables gradients to bypass multiple convolutional layers in the main path and propagate directly back to the input. This strategy avoids the issue of vanishing or exploding gradients, which can arise from the layer-by-layer multiplication of gradients during backpropagation through deep networks. By enabling the direct flow of gradients, the skip connection ensures that the model has sufficient parameter capacity to tackle a variety of complex tasks. In this paper, we employ the core concept of skip connections to provide our model with an adequate number of parameters. This allows the model to effectively learn and capture the complex correlations between single-pixel measurements and objects features. 

\subsubsection{The structure of Transformer2DModel}
As can be seen from Fig. \ref{fig2}(b), the Transformer2DModel primarily consists of GroupNorm layers, convolutional layers with a kernel size of 1×1, permute and reshape operations, fully connected layers, and multi-head attention mechanisms. Among these, the GroupNorm layers are used to normalize the extracted features, while the convolutional layers with a kernel size of 1×1 are used for task-relevant feature extraction. The permute operation transposes the dimensions of the features, and the reshape operation modifies the dimensions of the features. The fully connected layers integrate all previously extracted features, while the multi-head attention mechanism plays a central role in enabling Transformer model to learn spatial dependencies across pixels during the image reconstruction. As illustrated in Fig. \ref{fig2}(c), given an input, the multi-head attention mechanism first divides the input into i parts. These parts are then fed into fully connected layers to generate new feature maps: Query ($Q_i$), Key ($K_i$), Value ($V_i$). Subsequently, the model computes the spatial attention map by applying a softmax layer to the result of the matrix multiplication between the transpose of $Q_i$  and $K_i$ . This attention map encompasses the interdependencies between all feature points and the current feature point. During the image reconstruction, the attention mechanism allows the model to explore spatial dependencies between the current pixel and all other pixels, adjusting the current pixel's value based on these dependencies. Finally, the model carries out a matrix multiplication between the attention map and $V_i$ to obtain $O_i$, which is then concatenated to form the final feature. This process can be formulated as follows:

\begin{align}
O_i = \text{softmax}\left( \frac{Q_i^T \cdot K_i}{\sqrt{\frac{C}{H}}} \right) V_i
\end{align}

The multi-head attention mechanism within the Transformer2DModel enables the encoding of inputs from multiple perspectives, capturing diverse dependencies between different positions in the feature sequences. In this work, we employ this strategy to learn spatial dependencies across pixels during image reconstruction, aiding in the extraction of comprehensive object information for subsequent reconstruction steps. This significantly enhances the model’s capability to reconstruct complex objects.

\section{Simulation experiments}

To validate the effectiveness of GILM in reconstructing complex objects, we conducted simulation experiments using several complex binary and grayscale objects. The intensity sequences of these objects were simulated using Python. In the absence of specific declarations, the resolution of the simulated object image is 320×320, with an intensity sequence sampled 10,000 times and a sampling rate of $9.7\%$. We compared GILM with several other methods, including differential ghost imaging (DGI), GICS, CNN-based GI, and UNet-based GI. The CNN-based GI uses the network architecture designed by GIDL \cite{29}, while the UNet-based GI employs the network architecture designed by GIDC \cite{35}. To enhance the generalization ability of these models, both the CNN and UNet-based GI models were trained using a self-supervised approach. Additionally, to quantitatively assess the quality of the reconstructed images, we used Peak Signal-to-Noise Ratio (PSNR) and Structural Similarity Index (SSIM) as evaluation metrics. PSNR quantifies the difference between the reconstructed image and the real object, with higher values indicating less distortion and thus better image quality. SSIM, on the other hand, measures the similarity between the reconstructed image and the real object, where higher values reflect greater similarity to the original object.

\subsection{Image quality for simulated object with different methods}

\begin{figure*}[htbp]
  \centering
  \includegraphics[width=13cm]{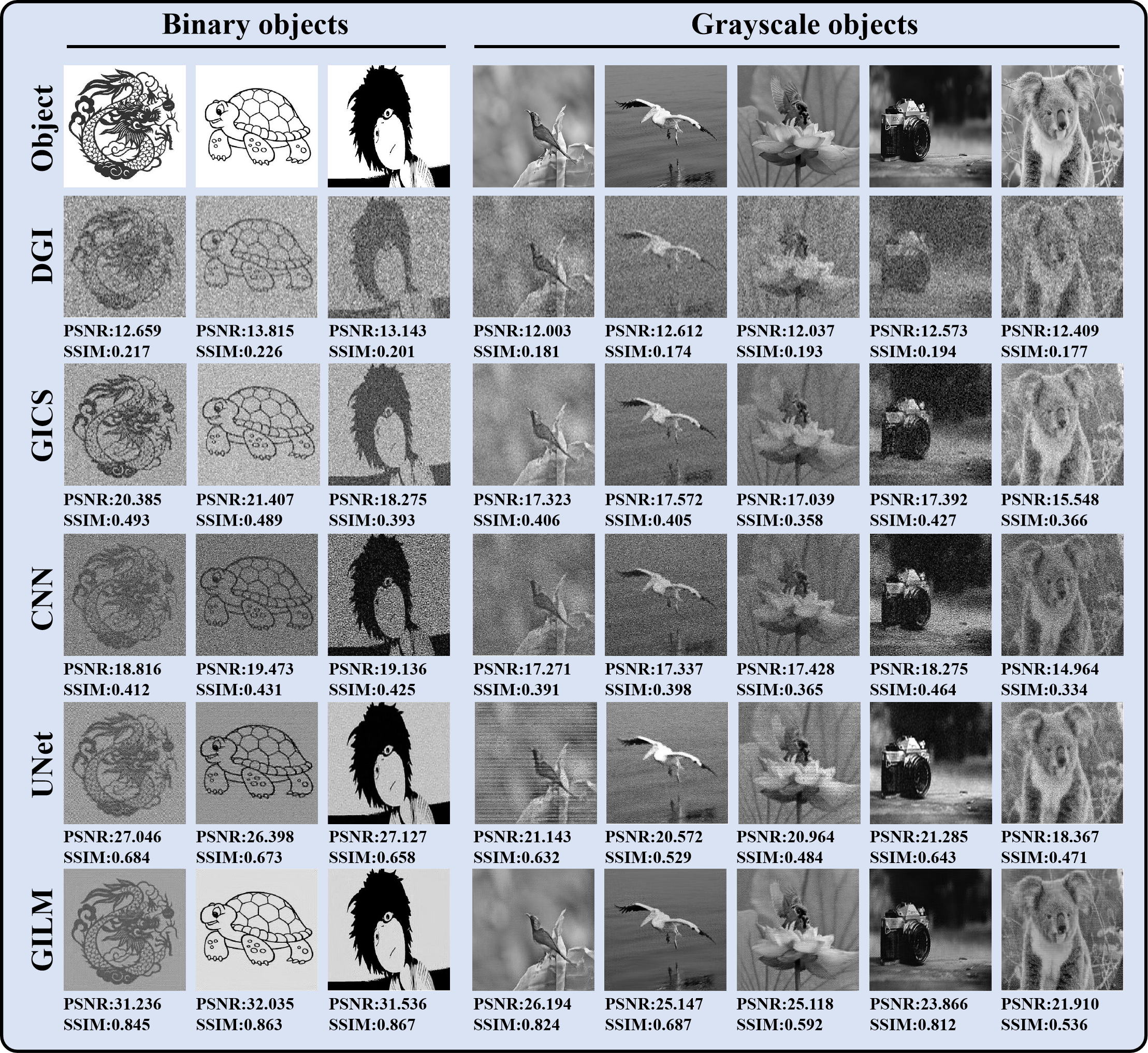}
  \caption{The results of retrieve the simulated binary and grayscale object images along with their corresponding PSNR and SSIM using the DGI, GICS, CNN-based GI, UNet-based GI, and GILM methods
 }
  \label{fig3}
\end{figure*}

We initially test the effectiveness of the proposed method on simulated binary objects. As shown in Fig. \ref{fig3}, the clarity of the DGI-reconstructed image, along with its corresponding PSNR and SSIM performance, is the worst. The performance progressively improves with GICS, CNN, and UNet, with GILM achieving the best results. This can be attributed to the fact that DGI reconstructs object images based on the linear correlation between collected light intensity signals and speckle patterns, which makes it difficult to achieve high-quality reconstruction of the object image at a sampling rate of $9.7\%$. In contrast, compressed sensing can recover sparse signals from a small amount of measurement data, extracting as much information as possible from incomplete samples. Therefore, at the same sampling rate, the image quality reconstructed by GICS outperforms that of DGI. Both CNN and UNet model the image reconstruction process as a complex higher-order function using convolutional layers, establishing a mapping between the input information and the reconstructed image. By analyzing the fluctuations in light intensity, these models learn the object information and effectively remove redundant data during the reconstruction process, thereby enhancing the reconstruction quality. However, due to limitations in model parameters and network architecture design, both CNN and UNet often suffer from underfitting when reconstructing complex object images. This makes it difficult for them to accurately capture the correspondence between single-pixel measurements and intricate object features, ultimately limiting the visibility of the reconstructed images. The large imaging model uses skip connections to address gradient explosion issues caused by increased network depth, ensuring that the model has sufficient parameters to capture the mapping relationship between single-pixel measurements and complex object images. Furthermore, the model incorporates multi-head attention mechanisms to capture spatial dependencies between pixels during the reconstruction, which enables the model to extract richer and more comprehensive object information. This approach significantly enhances the model’s performance in reconstructing complex object images, demonstrating its great potential for high quality image reconstruction.

To further verify these observations, we conducted experiments on simulated grayscale objects. The results show that GILM maintains superior performance even when reconstructing complex grayscale object images, further supporting the conclusion that GILM shows significant potential in advancing GI reconstruction of complex object imaging.

\subsection{Image quality for simulated object on different iterations and sample rates}

\begin{figure*}[htbp]
  \centering
  \includegraphics[width=13cm]{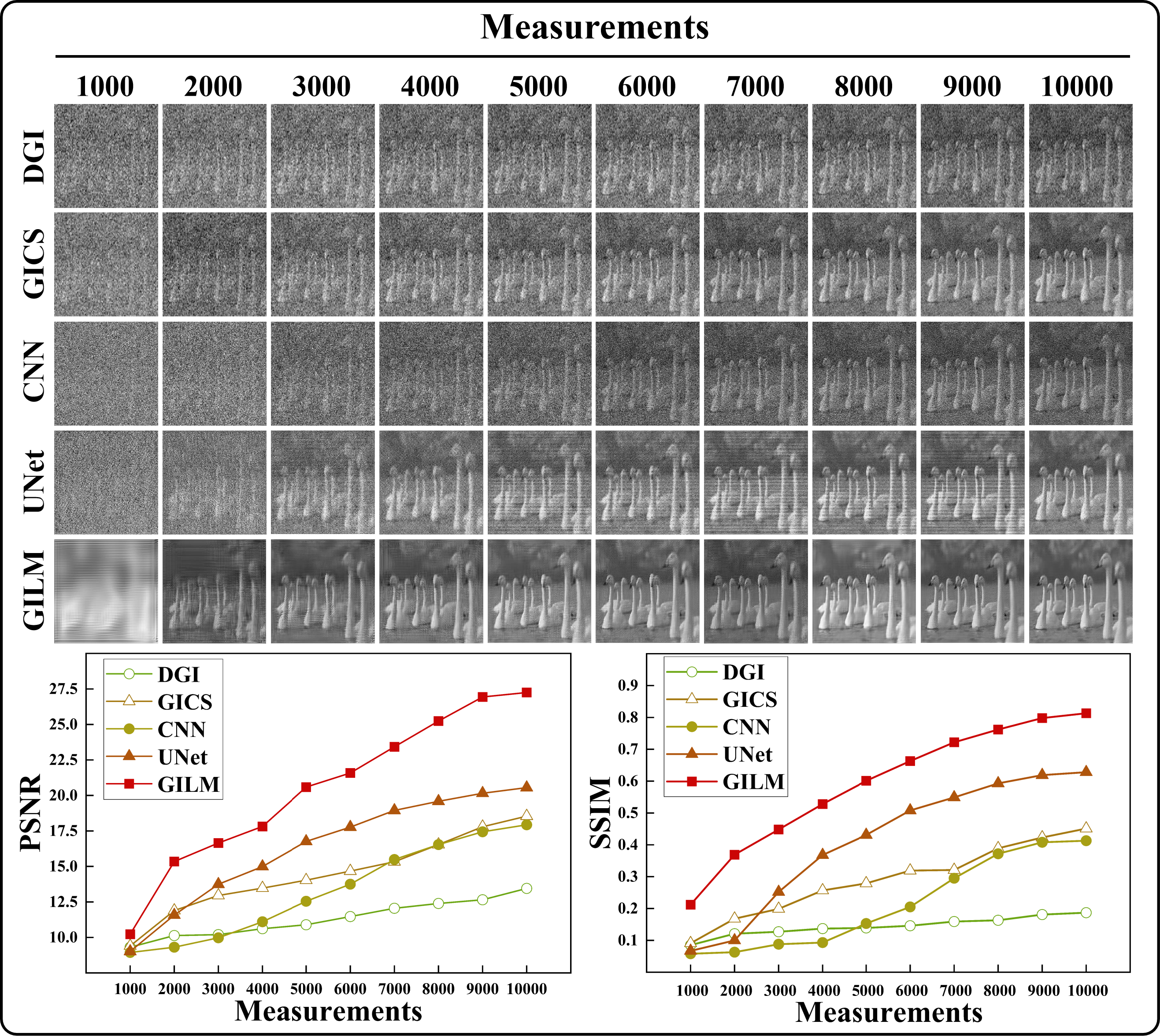}
  \caption{The results of retrieve the simulated grayscale object images at different measurements along with their corresponding PSNR and SSIM using the DGI, GICS, CNN-based GI, UNet-based GI, and GILM methods
 }
  \label{fig4}
\end{figure*} 

To further verify the effectiveness of GILM under different sample rates, we compared its performance with other GI methods across varying numbers of measurements, ranging from 1000 to 10000 at intervals of 1000. The corresponding sampling rates $\beta$ varied from $0.97\%$ to $9.7\%$ for an image resolution of 320×320, with each step increasing by $0.97\%$. As shown in Fig. \ref{fig4}, GILM demonstrates the highest fidelity in reconstructed images. Notably, GILM can reproduce the general outline of the target even at a $\beta$ as low as $1.94\%$, whereas UNet requires at least twice the measurements to achieve comparable results, and CNN, GICS and DGI require even more. This highlights the potential of GILM to achieve high-quality reconstruction of object images at low sampling rates.

To quantitatively evaluate the reconstructed images, we calculate the PSNR and SSIM for each reconstruction. The PSNR values are presented in the bottom-left part of Fig. \ref{fig4}, while the SSIM values are displayed in the bottom-right. It is evident that as the number of measurements increases, the PSNR and SSIM values for images reconstructed by DGI and GICS increase slowly. This is primarily due to that DGI and GICS reconstructs the object image through linear calculations of speckle patterns and collected light intensity, which is less efficient in information utilization, thus necessitating a larger number of measurements for high quality reconstruction. In contrast, CNN, UNet, and GILM leverage the superior feature extraction capabilities of DL, enabling them to make more efficient use of the collected data. Consequently, these models show a rapid improvement in the PSNR and SSIM of the reconstructed images as the measurement increases. It is clear that GILM consistently achieves the highest metrics across all scenarios, further validating its capability in reconstructing object images with high fidelity.

\begin{figure*}[htbp]
  \centering
  \includegraphics[width=12cm]{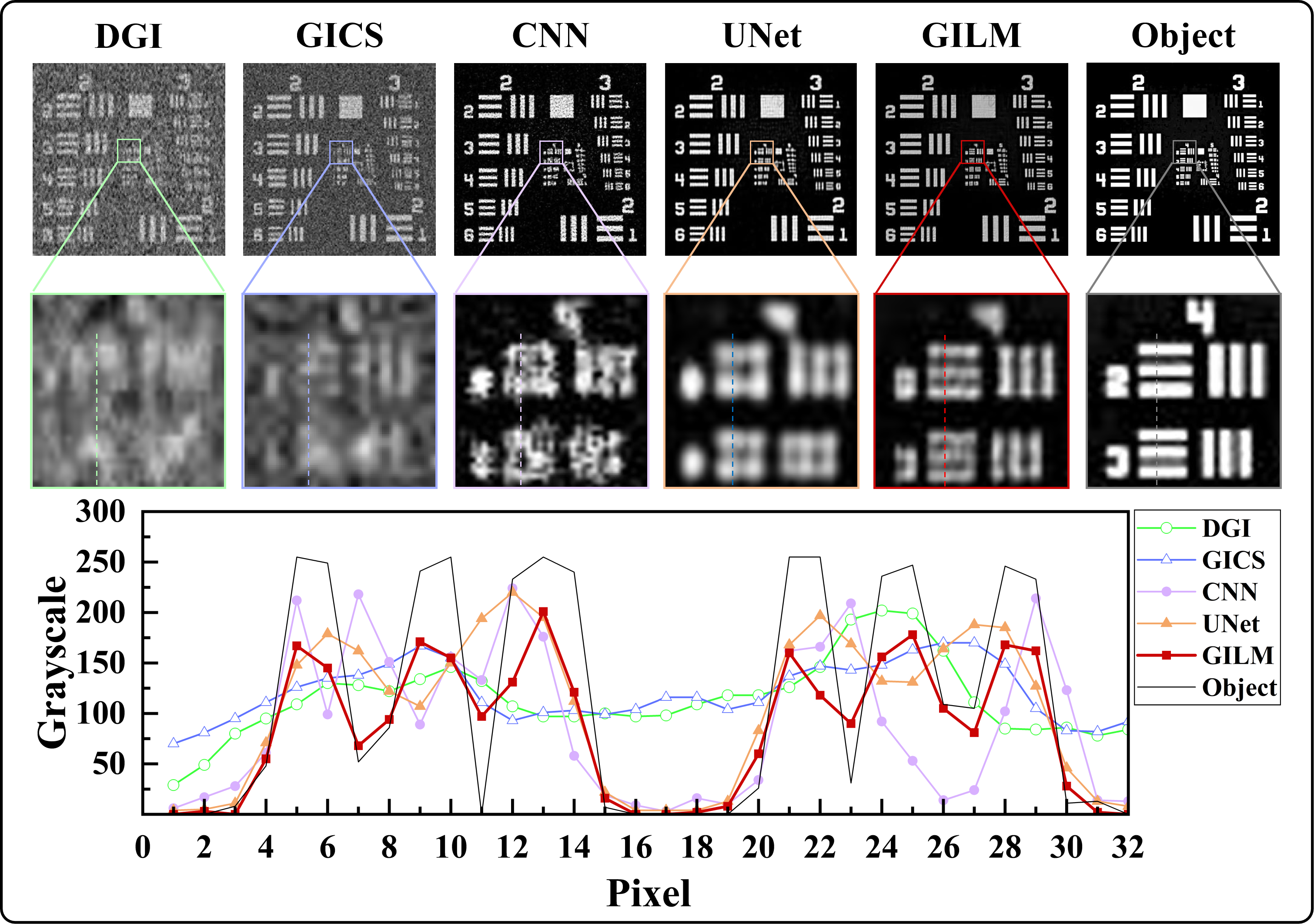}
  \caption{The experiment results of retrieve the USAF resolution target along with their grayscale distribution at the mark using the DGI, GICS, CNN-based GI, UNet-based GI, and GILM methods
 }
  \label{fig5}
\end{figure*} 

\subsection{Image quality for simulated object at different image size}

To validate the effectiveness of GILM in reconstructing object details, we compared its performance with other GI methods to reconstruct USAF resolution target with 320×320 pixels. We also proceed to reconstruct object images of various sizes to explore GILM’s potential for high-resolution imaging. As illustrated in Fig. \ref{fig5}, the images reconstructed using DGI and GICS display intense background noise, while those reconstructed with CNN, UNet, and GILM do not exhibit noticeable background noise. This is largely attributed to DL-based GI algorithms model high-order functions for nonlinear analysis of collected signals, leading to a notable enhancement in image quality at the same sampling conditions. To further assess GILM’s ability to reconstruct fine image details, we examine the grayscale distribution in marked regions of the reconstructed images. The results show that GILM can effectively distinguish the object and background areas even in very detailed areas, while other GI methods tend to misidentify the background as part of the object. This is primarily because GILM leverages the multi-head attention mechanism to learn spatial dependencies across pixel points during image reconstruction. This allows the model to extract comprehensive object information, significantly improving the clarity and accuracy of the reconstructed images.

\begin{figure*}[htbp]
  \centering
  \includegraphics[width=13cm]{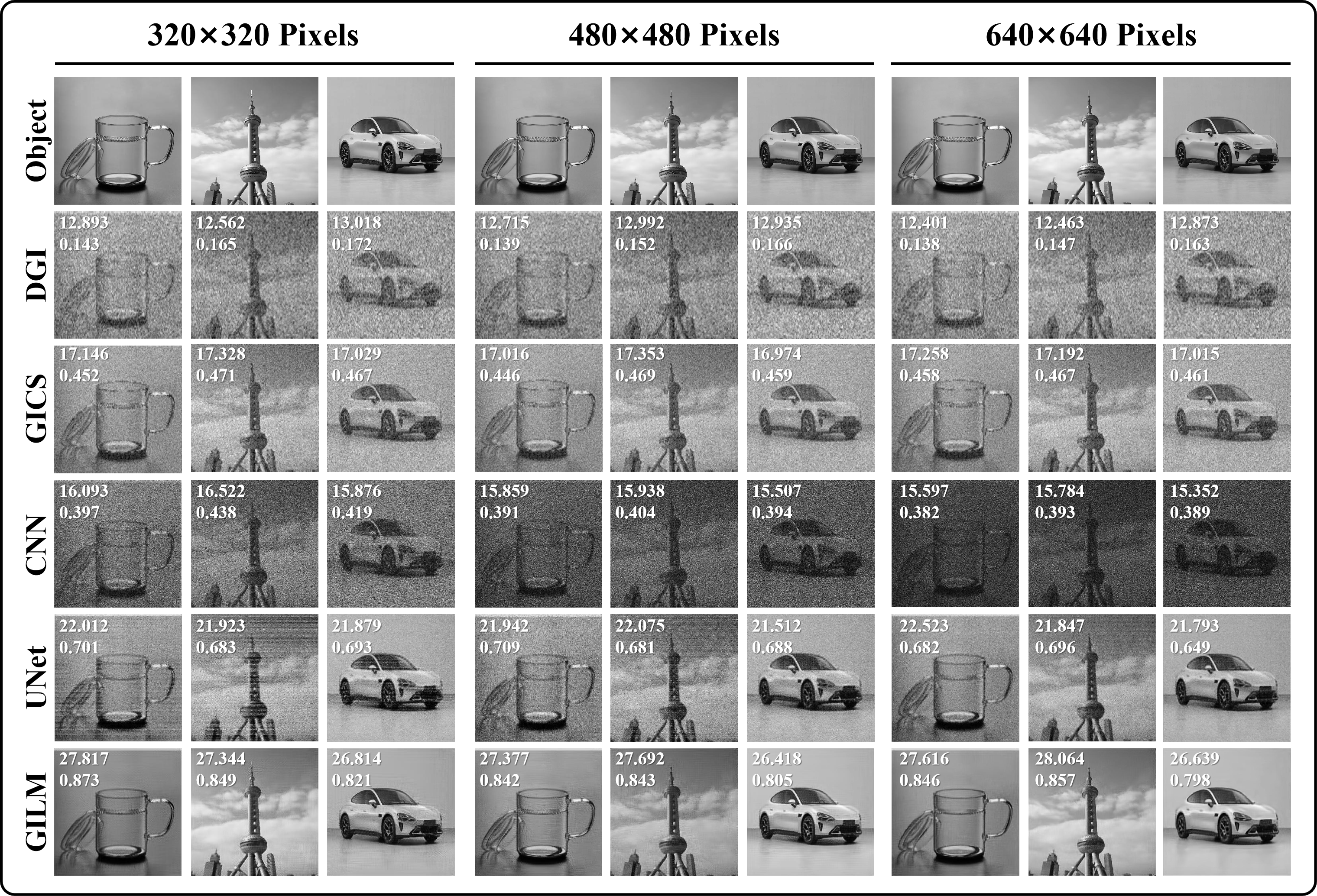}
  \caption{The results of retrieve the simulated grayscale object images at different resolutions along with their corresponding PSNR and SSIM using the DGI, GICS, CNN-based GI, UNet-based GI, and GILM methods 
 }
  \label{fig6}
\end{figure*}

After verifying GILM’s ability to reconstruct object details, we further investigate the potential of GILM in reconstructing object images with high resolution. The resolution of reconstructed images progressively increases from 320×320 to 480×480, and then to 640×640 pixels. For each case, the number of measurements was kept constant at 10000, resulting in corresponding sampling ratios are $9.7\%$, $4.3\%$, and $2.4\%$, respectively. As demonstrated in Fig. \ref{fig6}, GILM successfully reconstructs object images with high fidelity regardless of the resolution, an assertion supported by the PSNR and SSIM evaluations. Notably, while increasing the resolution leads to a linear decrease in the sampling ratio when the measurements remain fixed, there is no marked deterioration in the quality of images reconstructed by GILM, even at the highest resolution of 640×640 pixels. This evidence highlights the robust potential of GILM in the domain of high-resolution GI.

\section{Experiments in free space and underwater environments}

\subsection{Free space GI experiment}

\begin{figure*}[htbp]
  \centering
  \includegraphics[width=10.5cm]{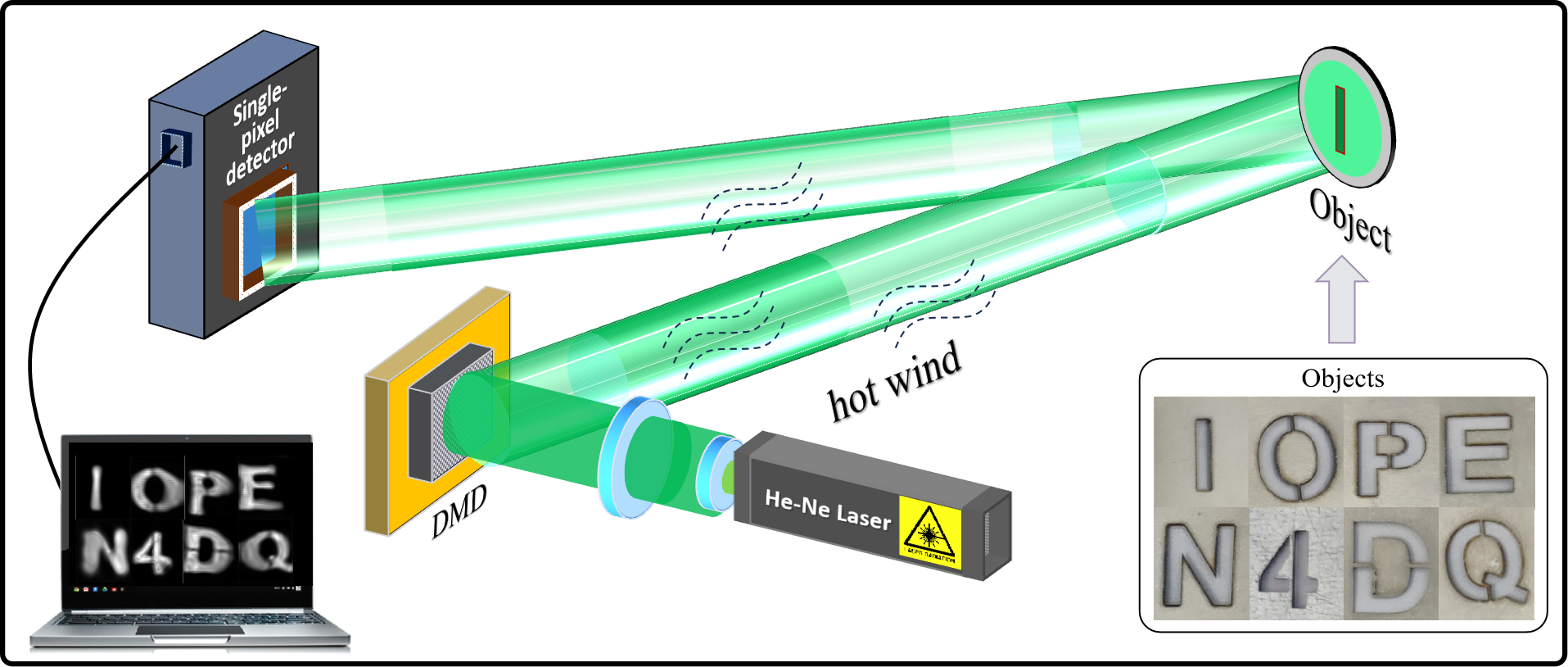}
  \caption{The specific experimental setup of free space GI experiments
 }
  \label{fig7}
\end{figure*}

\begin{figure*}[htbp]
  \centering
  \includegraphics[width=10.5cm]{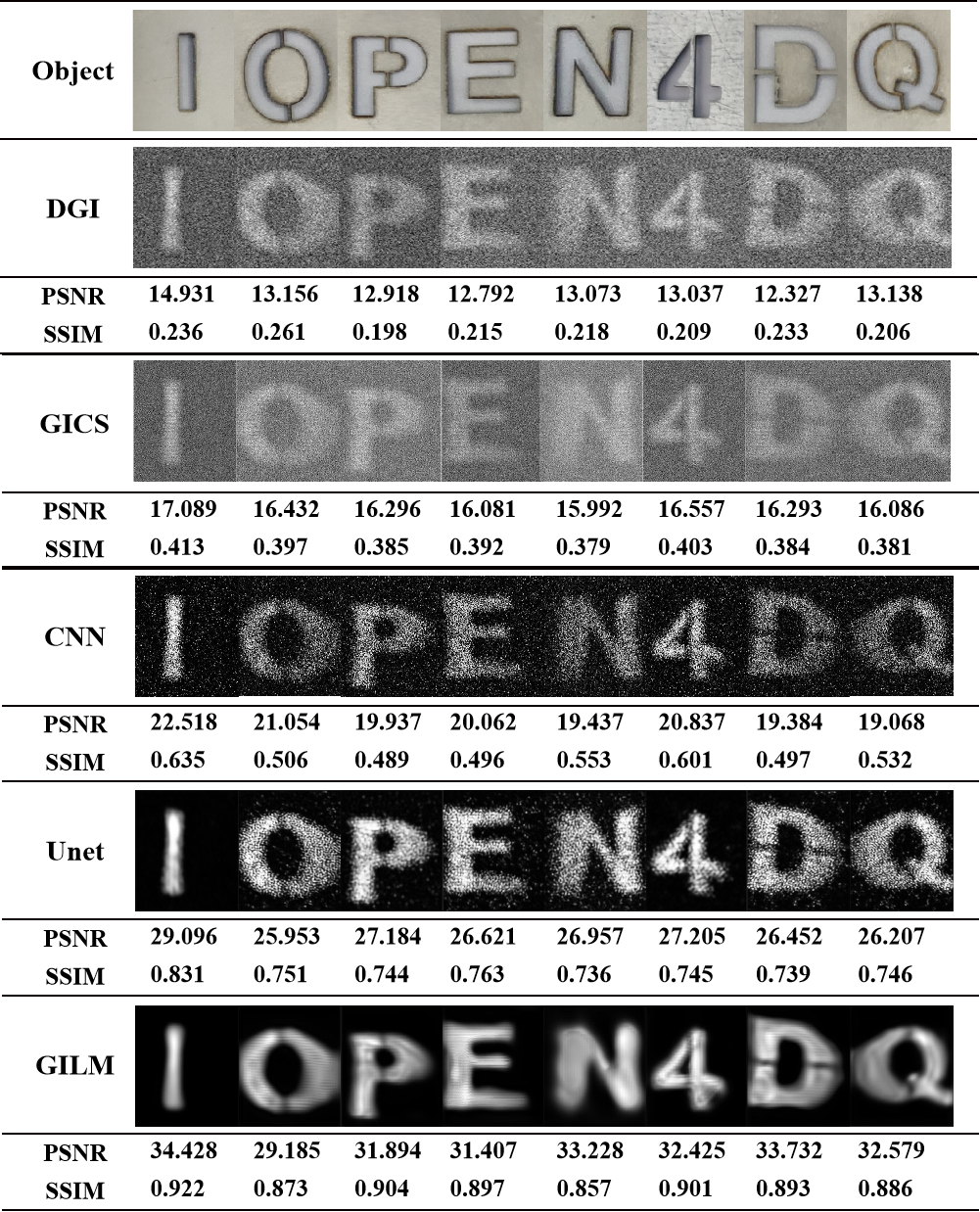}
  \caption{The results of retrieve the true object images in free space along with their corresponding PSNR and SSIM using the DGI, GISC, CNN-based GI, UNet-based GI, and GILM methods
 }
  \label{fig8}
\end{figure*}

To further validate the efficacy of GILM in real-world applications, we perform image reconstruction experiments on eight letter/number targets: “I”, “O”, “P”, “E”, “N”, “4”, “D”, and “Q” in free space. The specific experimental setup of free space GI experiments are shown in Fig. \ref{fig7}. Specially, a laser is used as light source, a digital micromirror device (DMD) is used to modulate the light beam emitted by the laser source, the SPD is used to receive the 1D temporal intensity signal of light passing through the object. Finally, the collected signals are transmitted to the computer for image reconstruction. In addition, to validate the proposed method’s robustness against scattering effects, we conducted turbulence simulations using heated airflow. The experiment results, as illustrated in Fig. \ref{fig8}, indicate that the images reconstructed by GILM exhibit the strongest object signal and the least background noise. The quantitative PSNR and SSIM indicators also suggest that the images reconstructed by GILM most closely resemble the actual objects. Notably, all reconstruction methods exhibit minor image distortions attributable to light field perturbations under turbulent conditions. These findings substantiate that GILM effectively analyzes the fluctuations in the acquired signals, thus maximizing the recovery of object images from collected signals.

The experimental results in Figs. \ref{fig3}-\ref{fig6} and Fig. \ref{fig8} clearly demonstrate that, when imaging the USAF resolution target and true objects in free space, the performance of CNN-based GI significantly outperforms GICS. However, when imaging simulated complex binary and grayscale objects, the performance of CNN is either comparable to or even worse than that of GICS. This discrepancy is primarily attributed to the limitations of CNN in terms of model parameters and network architecture, which often lead to underfitting when imaging complex objects. Consequently, this underfitting hampers the establishment of the mapping relationship between the input information and the reconstructed image, thereby degrading the quality of the reconstructed images. This phenomenon also highlights the significance of introducing large model technology into the field of computational imaging.

\subsection{Underwater GI experiment}

\begin{figure*}[htbp]
  \centering
  \includegraphics[width=13cm]{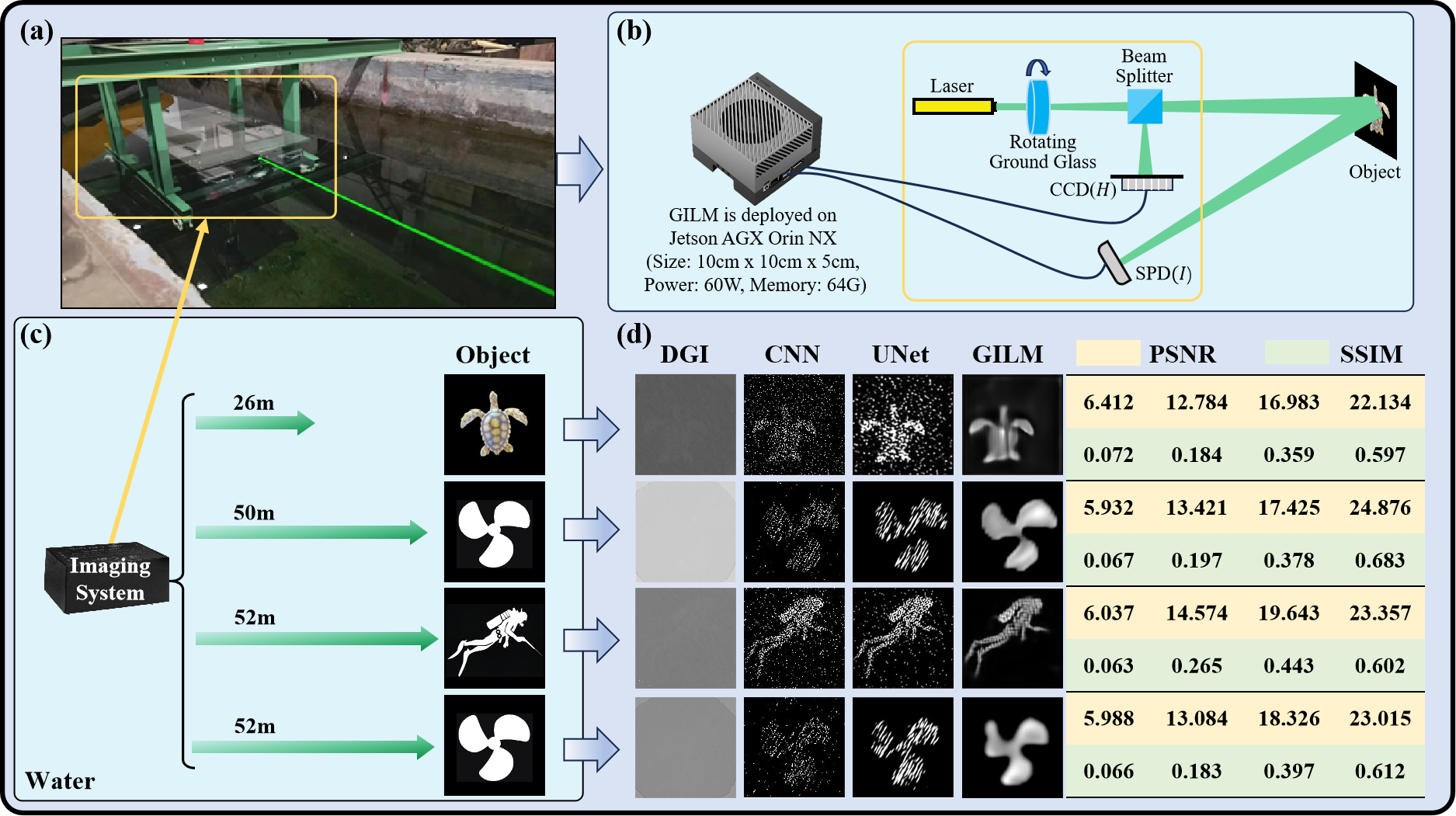}
  \caption{Overview of underwater GI experiment. (a) the environment of the underwater GI experiment, (b) the specific experimental setup of underwater GI experiments, (c) the underwater imaging objects and their distances, (d) the images reconstructed by DGI, CNN-based GI, UNet-based GI, and GILM, along with the respective PSNR and SSIM metrics, where the first line is PSNR and the second line is SSIM.
 }
  \label{fig9}
\end{figure*} 

To further validate the capability of GILM in imaging underwater objects, we conduct image reconstruction experiments in an underwater environment. The full framework of the underwater ghost imaging experiment is depicted in Fig. \ref{fig9}. Figure \ref{fig9}(a) shows the experimental environment. Figure \ref{fig9}(b) provides specific experimental equipment. Specifically, a laser is used as the light source, with a rotating ground glass to modulate the light field. During the propagation of light, a beam splitter divides the light into two paths: one path is transmitted to the object, and the reflected 1D light intensity signal is collected by the SPD; the other path is transmitted to the charge coupled device (CCD), which is used to record the 2D light field information. From Fig.\ref{fig9}(c), it can be observed that the objects in this experiment consist of a turtle at 26 meters, a propeller at 50 meters, and another propeller and a frogman both situated at 52 meters away. The image reconstruction outcomes for these objects, together with the corresponding PSNR and SSIM metrics, are displayed in Fig. \ref{fig9}(d). The results indicate that the quality of image reconstructed by GILM significantly outperforms other GI methods, a fact also substantiated by the quantitative PSNR and SSIM indicators. The underwater GI experiment further confirms GILM’s ability to reconstruct complex object images. Additionally, when the imaging resolution is set to 320×320, 480×480, and 640×640, running GILM requires 29GB, 34GB, and 39GB of GPU memory, respectively. We have successfully deployed it on a portable computing platform (Jetson AGX Orin NX) with 64GB of memory, demonstrating its feasibility for practical engineering applications.

\section{Conclusion}
In this study, we propose a large imaging model with 1.4 billion parameters, integrating the physical model of GI into the proposed large imaging model. GILM employs skip connection mechanisms to mitigate the inherent gradient explosion challenges of deep architectures, ensuring sufficient parameter capacity to capture the complex correlations between single-pixel measurements of objects. Furthermore, GILM utilizes multi-head attention mechanisms to learn spatial dependencies across pixel points during the image reconstruction process, which aids in extracting comprehensive object information for subsequent reconstruction. We conduct a series of imaging experiments to validate the effectiveness of GILM, including simulated object imaging, imaging objects in free space, and imaging object located 52 meters away in underwater environment. The experimental results indicate that GILM effectively analyzes the fluctuation trends of the collected signals, thereby maximizing the recovery of object images from the acquired signals. It is worth mentioning that we have successfully deployed GILM on a portable computing platform, demonstrating its feasibility for practical engineering deployment. In the future, we aim to introduce the GILM algorithm into more computational optical imaging fields, such as phase imaging and polarization imaging, to enhance visual applications by improving image reconstruction quality and enabling more accurate and detailed visual information extraction. The source code and dataset can be found at https://github.com/BestAnHongjun/GILM.

\section{Authors’ contributions} 
Yifan Chen conceived the initial idea, performed the experiments and analyzed the data. Hongjun An  participated in discussions and performed the experiments. Zhe Sun  analyzed the data and supervised the project. Tong Tian participated in discussions. Mingliang Chen conducted the underwater experiment. Christian Spielmann participated in discussions. Xuelong Li supervised the project.

\section{Funding} 
This work was supported by the Natural Science Basic Research Program of Shaanxi (No. 2024JC-YBMS-468), the State Key Laboratory for Underwater Information and Control (No. 2024-CXPT-GF-JJ-036-09), and the National Key R$\&$D Program of China (No. 2022YFC2808003)

\section{Data availability} 
The datasets used and analysed during the current study are available from the corresponding author on reasonable request.

\section{Conflict of interest} 
The authors declare no competing interests.


\end{document}